\documentclass{PoS}

\usepackage{amssymb,amsmath,graphicx}

\newcommand{\SLNC}[1]{{\rm SL}(#1, \mathbb{C})}
\newcommand{\SUN}[1]{{\rm SU}(N)}

\newcommand {\beq} {\begin{equation}}
\newcommand {\eeq} {\end{equation}}
\newcommand {\beqa}{\begin{eqnarray}}
\newcommand {\eeqa}{\end{eqnarray}}

\newcommand{\1}{\mbox{1}\hspace{-0.25em}\mbox{l}}

\newcommand{\calU}{{ U}}  
\newcommand{\calN}{{\cal N}}
\newcommand{\eps}{\epsilon}

\title{Gauge cooling for the singular-drift problem in the complex Langevin method --- an application to finite density QCD}

\ShortTitle{Gauge cooling for the singular-drift problem in the CLM - an application to FDQCD}

\author{\speaker{Keitaro Nagata}$^a$\thanks{KEK-TH1941}, Hideo Matsufuru$^{a}$, 
Jun Nishimura$^{a,b}$, Shinji Shimasaki$^{c}$, \\
$^{a}$ KEK , High Energy Accelerator Research Organization, 1-1 Oho, Tsukuba, Ibaraki 305-0801, Japan \\
$^{b}$ Graduate University for Advanced Studies (SOKENDAI),
1-1 Oho, Tsukuba, Ibaraki 305-0801, Japan \\
$^{c}$ Research and Education Center for Natural Sciences, Keio University,
Hiyoshi 4-1-1, Yokohama, Kanagawa 223-8521, Japan\\
E-mail: \email{matufuru@post.kek.jp,knagata@post.kek.jp,jnishi@post.kek.jp,\\
shinji.shimasaki@keio.jp}
}

\abstract{We study full QCD at finite density and low temperature 
with light quark mass using the complex Langevin method. 
Since the singular drift problem turns out to be mild
on a $4^3 \times 8$ lattice we use,
the gauge cooling is performed only to control the unitarity norm 
in this exploratory study.
We report on our preliminary data obtained 
from the complex Langevin simulation up to certain Langevin time.
While the data are still noisy due to lack of statistics, 
the onset of the baryon number density seems to occur
at larger $\mu$ than half the pion mass, which is
the value for the phase quenched QCD. 
The validity of our simulation is tested by 
the recently proposed criterion based on
the probability distribution of the drift term.}

\FullConference{34th Annual International Symposium on Lattice Field Theory\\
                24-30 July 2016\\
                University of Southampton, UK}

\begin{document}

\section{Introduction}

In recent years we have witnessed 
remarkable progress in the study of systems 
with complex actions by the complex Langevin method (CLM).
The CLM is based on the complexification of dynamical variables,
which requires some conditions for the equivalence to the original theory.
Although the conditions had not been known for a long time, 
Ref.~\cite{Aarts:2009uq,Aarts:2011ax}
provided an explicit condition based on an argument for 
justification of the CLM.
In order to satisfy this condition in the case of lattice QCD,
the so-called gauge cooling was developed \cite{Seiler:2012wz}.
Thus, the complex Langevin simulation for QCD was realized 
in the QGP phase~\cite{Sexty:2013ica,Fodor:2015doa}
and in the heavy dense limit~\cite{Aarts:2016nju} 
up to quite large quark chemical potential,
which goes far beyond the applicable limit of conventional approaches 
such as reweighting or the Taylor expansion.
The validity of the gauge cooling was also proved 
explicitly \cite{Nagata:2015uga}
based on the argument for justification of the CLM.
See also other contributions to this volume for recent studies on the 
CLM~\cite{Sinclair:2016nbg,Schmalzbauer:2016pbg,Aarts:2016mso,Attanasio:2016mhc}.

In this work, we apply the CLM to QCD in the confined phase with light quarks.
In this case, it is anticipated that the appearance of zero eigenvalues 
of the fermion matrix makes the drift term singular and
spoils the validity of the CLM.
This problem was shown to occur 
in some models \cite{Mollgaard:2013qra,Greensite:2014cxa,Nishimura:2015pba},
and possible solutions 
were
discussed \cite{Mollgaard:2014mga,Nagata:2015ijn,Nagata:2016alq}.
In this work, we perform simulations on a $4^3 \times 8$ lattice,
where the singular drift problem turns out to be mild.
For this reason, we use
the gauge cooling only to control the unitarity norm
instead of using it also to cure 
the singular drift problem as proposed 
in Refs.~\cite{Nagata:2015ijn,Nagata:2016alq}.
The validity of our simulation is tested
by using a criterion that
the probability distribution of the drift term 
should fall off exponentially or faster \cite{Nagata:2016vkn},
which is required for the convergence property of
a power series used in a refined argument for justification.
This condition is considered a necessary and sufficient
condition since it is slightly stronger than
the previous condition required for the validity 
of the integration by parts \cite{Aarts:2009uq,Aarts:2011ax}.
The newly proposed criterion is shown to be useful
not only in simple one-variable models \cite{Nagata:2016vkn}
but also in systems with infinite degrees 
of freedom~\cite{Shimasaki:2016lat,Ito:2016efb}.


We explain the framework of this work in the next section, 
provide preliminary results in section \ref{sec:result}, and 
draw a temporal conclusion in the final section. 

\section{Framework}
\label{sec:framework}


We study QCD on a four-dimensional Euclidean lattice with
the lattice spacing $a$, the spatial extent $N_s$ and 
the temporal extent $N_t$.
Introducing quark chemical potential $\mu$ makes 
the fermion determinant complex and spoils the importance sampling. 
We use the CLM to solve this problem.
When one applies the Langevin method to a theory with a complex action, 
the dynamical variables in the theory have to be complexified.
In lattice QCD, in particular, 
the space on which the link variables $U_{n\mu}$ 
take values should be extended from SU(3) to $\SLNC{3}$.
Accordingly, we need to extend the action to a function of 
the complexified link variables in a holomorphic manner. 
In this work, we use the plaquette action
\begin{align}
S_{\rm G} & = - \frac{\beta}{6} \sum_{x} \sum_{\mu > \nu} \,
{\rm tr} [ U_{x, \mu \nu} + U^{-1}_{x, \mu \nu}] 
\label{eq:2016Aug14eq1}
\end{align}
for the gauge field, where $U_{x\mu}\in \SLNC{3}$ is a link variable and 
$U_{x, \mu\nu}$ is the plaquette given by 
$U_{x, \mu\nu} = 
U_{x \mu} U_{x+\hat{\mu}\, \nu} U_{x+\hat{\nu} \, \mu}^{-1} U_{x\nu}^{-1}.$
Fermions are implemented using the standard staggered fermion formalism
with the fermion matrix
\begin{align}
M(U, \mu)_{xy} = m a\,  \delta_{x,y} + \sum_{\nu=1}^{4} \frac{\eta_\nu(x)}{2} 
\left[ e^{ \mu a \delta_{\nu 4}} U_{x\nu} \delta_{x+\hat{\nu},y} 
- e^{- \mu a \delta_{\nu 4}} 
U_{x-\hat{\nu}\nu}^{-1}\delta_{x-\hat{\nu},y}\right] \ ,
\end{align}
which represents four flavors of quarks with
the mass $m$ and the chemical potential $\mu$.
Due to the relation
$\epsilon_x M(\mu)_{x y} \epsilon_y = M^\dagger (-\mu^*)_{yx}$ 
with the ``staggered $\gamma_{\, 5}$ operator'' $\epsilon_{x}$, 
the fermion determinant becomes complex for real and nonzero $\mu$.
As interesting observables, 
we measure the baryon number density 
$\langle n \rangle = \frac{1}{N_V N_c} \frac{ \partial}{\partial (\mu a)} \log Z$
and the chiral condensate 
$\Sigma = \frac{1}{N_V} \frac{ \partial}{\partial (m a)} \log Z,$
where $N_V=N_s^3 N_t$.
In measuring these observables in the CLM, they should also be 
extended to functions of complexified 
link variables in a holomorphic manner.

In the CLM including the gauge cooling procedure, the link 
variables are updated in the following two steps 
\begin{align}
\widetilde{\calU}_{x\nu}(t) &= g_x \calU_{x\nu}(t) g_{x+\hat{\nu}}^{-1} \ , 
\label{eq:gauge-cooling}\\
\calU_{x\nu}(t+\eps) &= \exp\left(i \sum_{a=1}^{N_c^2-1} 
\lambda_a \left[  - v_{ax\nu}(\widetilde{U}) \epsilon + \sqrt{ 2 \epsilon} \, 
\eta_{a x \nu}\right]  \right) \widetilde{\calU}_{x\nu}(t) \ ,
\label{eq:cLangevineq}
\end{align}
where $t$ is the discretized Langevin time with the stepsize $\eps$
and $\lambda_a$ are the Gell-Mann matrices.
Eq.~(\ref{eq:cLangevineq}) 
represents the Langevin equation for the link variables.
The Gaussian white noise $\eta_{a x \nu}$ is real and 
normalized by $\langle \eta_{a' x' \nu'}(t')\eta_{a x \nu}(t)\rangle_\eta
= 2 \delta_{a'a}\delta_{x'x}\delta_{\nu'\nu}\delta_{t't}$,
where the symbol $\langle ... \rangle_\eta$ represents
an average over the noise with the Gaussian weight.
The drift term is defined by 
\begin{align}
v_{a x\nu}(U) = 
\lim_{\delta\to 0} 
\frac{ S( e^{ i \delta \lambda_a} \calU_{x\nu}) - 
S( \calU_{x\nu})}{\delta} \ , 
\end{align}
where $S= S_G - \frac{N_f}{4} \log \det M$ with $N_f=4$ in our case.
At $\mu \neq 0$, the action $S$ is complex, and so is $v_{ax\nu}$.
Therefore, the use of Eq.~(\ref{eq:cLangevineq}) makes
the link variables drift away from the SU(3) manifold
in the non-compact directions.

In calculating the fermionic part of the drift term, 
we use the bilinear noise method, where the trace 
of $M$ is evaluated using the Gaussian noise vectors. 
It is known 
that the use of this method in the Langevin simulation
does not give rise to systematic errors 
in the zero stepsize limit \cite{Batrouni:1985jn},
which is in contrast to the situation in the hybrid R algorithm.
While a naive implementation of this method violates 
the reality of the drift term at $\mu=0$, 
there is an improved implementation that makes the drift term real 
at $\mu=0$ \cite{Sinclair:2015kva}, which we follow in our study.

Due to the complexification of dynamical variables, 
the gauge invariance of the action and 
the observables is enhanced from SU(3) to $\SLNC{3}$.  
The gauge cooling represented by Eq.~(\ref{eq:gauge-cooling}) 
is performed by using a transformation for this enhanced symmetry.
%
%
We determine $g_x$ in Eq.~(\ref{eq:gauge-cooling}) 
in such a way that 
it reduces the unitarity norm \cite{Sexty:2013ica,Fodor:2015doa}
\begin{align}
\calN_{\rm u} \equiv 
\frac{1}{4 N_V}\sum_{x,\nu} {\rm tr} 
[ (\calU_{x\nu})^\dagger \calU_{x\nu} + 
(\calU^{-1}_{x\nu})^\dagger \calU^{-1}_{x\nu} -2 \times \1_{3\times 3} ] \ , 
\label{eq:unitarity-norm}
\end{align}
which describes the deviation of ${\calU}_{x\nu}$ from SU(3) matrices.
Note that 
the Hermitian conjugate in $\calN_{\rm u}$ is taken after the complexification
and that the unitarity norm is not invariant 
under an $\SLNC{3}$ transformation.

In order to test the validity of our simulation,
we calculate the probability of the drift term
\begin{align}
p(u) &=  \frac{1}{4 N_V}\Big\langle \sum_{x\nu} 
\delta\left(u-u_{x\nu}\right) \Big\rangle
 \ , 
\end{align}
where $u_{x \nu}$ is the norm of the drift term 
for a link variable ${\calU}_{x \nu}$ defined by 
\begin{align}
u_{x\nu} &=  
\sqrt{ \frac{1}{N_c^2-1} \sum_{a=1}^{N_c^2-1} |v_{ax\nu}|^2 } \ .
\label{eq:norm-drift}
\end{align}
Since the drift term $v_{ax\nu}$ is calculated anyway in the CLM,
the calculation of $p(u)$ does not require significant additional cost.
In order for the CLM to be valid,
the probability $p(u)$ should fall off exponentially or faster
at large $u$ \cite{Nagata:2016vkn}.
Note that the drift term $v_{ax\nu}$ 
contains the gauge field part and the fermionic part.
A slow fall-off of each part represents
the excursion problem and the singular drift problem, respectively.
Thus the two problems that were thought to invalidate the CLM
can be probed by the above criterion in a unified manner.
Without this criterion, it is not obvious to tell whether
either of these problems is really occurring or not.

\section{Results}
\label{sec:result}

\begin{figure}[htbp] 
\begin{center}
\includegraphics[width=7cm]{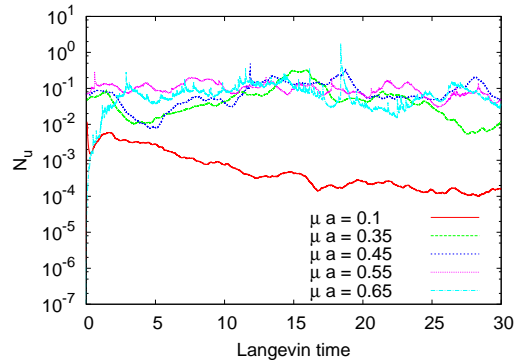}
\caption{The Langevin time history of the unitarity norm 
for $0.1 \le \mu a  \le  0.65$.}
\label{Fig:unorm_history}
\end{center}
\end{figure}

The parameters of the theory is chosen as
$N_s^3\times N_t=4^3 \times 8$, $m a = 0.05$ and $\beta = 5.7$, 
which corresponds to the confined phase at $\mu=0$. 
The chemical potential $\mu$ is varied in the range $\mu a\in [0, 0.65]$, 
which corresponds to $\mu/T \in [0, 5.2]$.
The Langevin simulations were performed with 
a fixed stepsize $\eps = 10^{-4}$ for the total Langevin time $t=30$.
For the sake of comparison, we also study the phase quenched QCD,
which is defined by replacing $\det M$ with $|\det M|$, 
using the (real) Langevin simulation 
with a stepsize $\eps = 2 \times 10^{-4}$ 
for the total Langevin time $t=20$.
The Langevin time history of the unitarity norm 
is plotted in Fig.~\ref{Fig:unorm_history}, which shows that
the unitarity norm is under control during the simulations.

\begin{figure}[htbp] 
\begin{center}
\includegraphics[width=7cm]{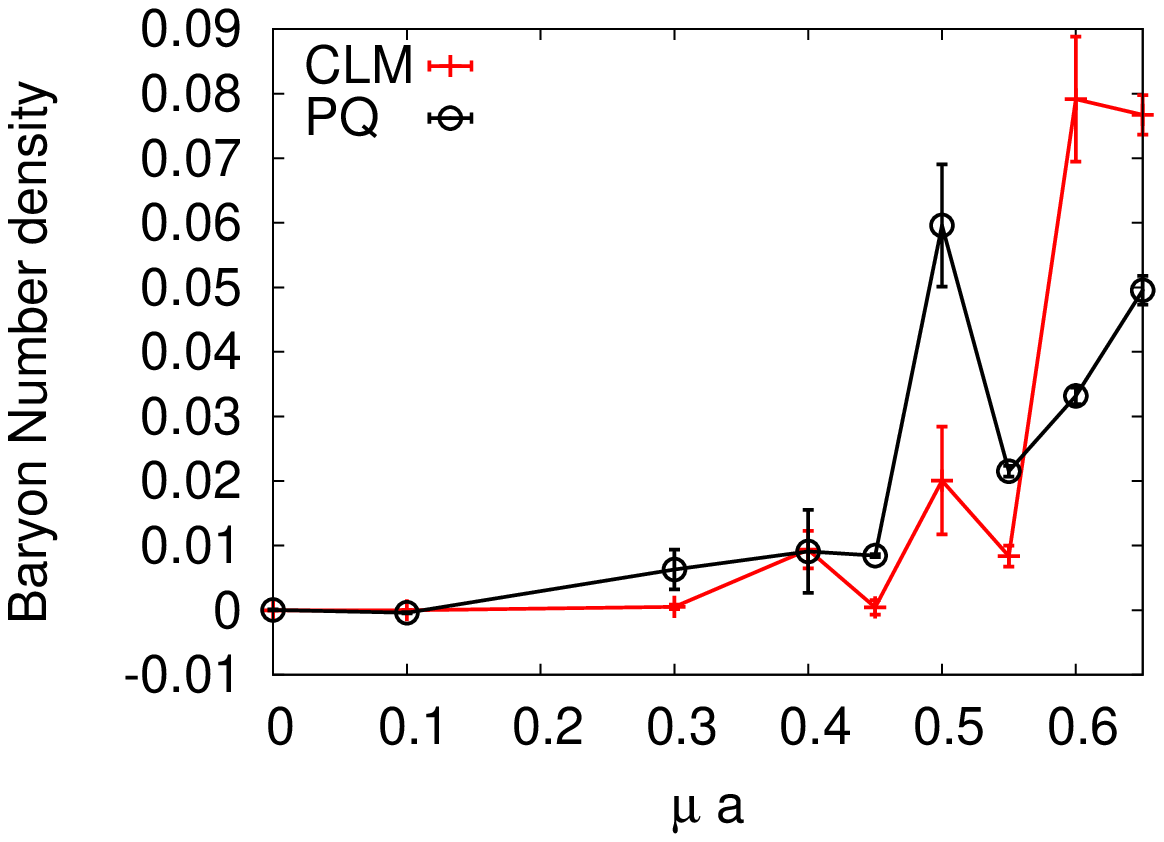}
\includegraphics[width=7cm]{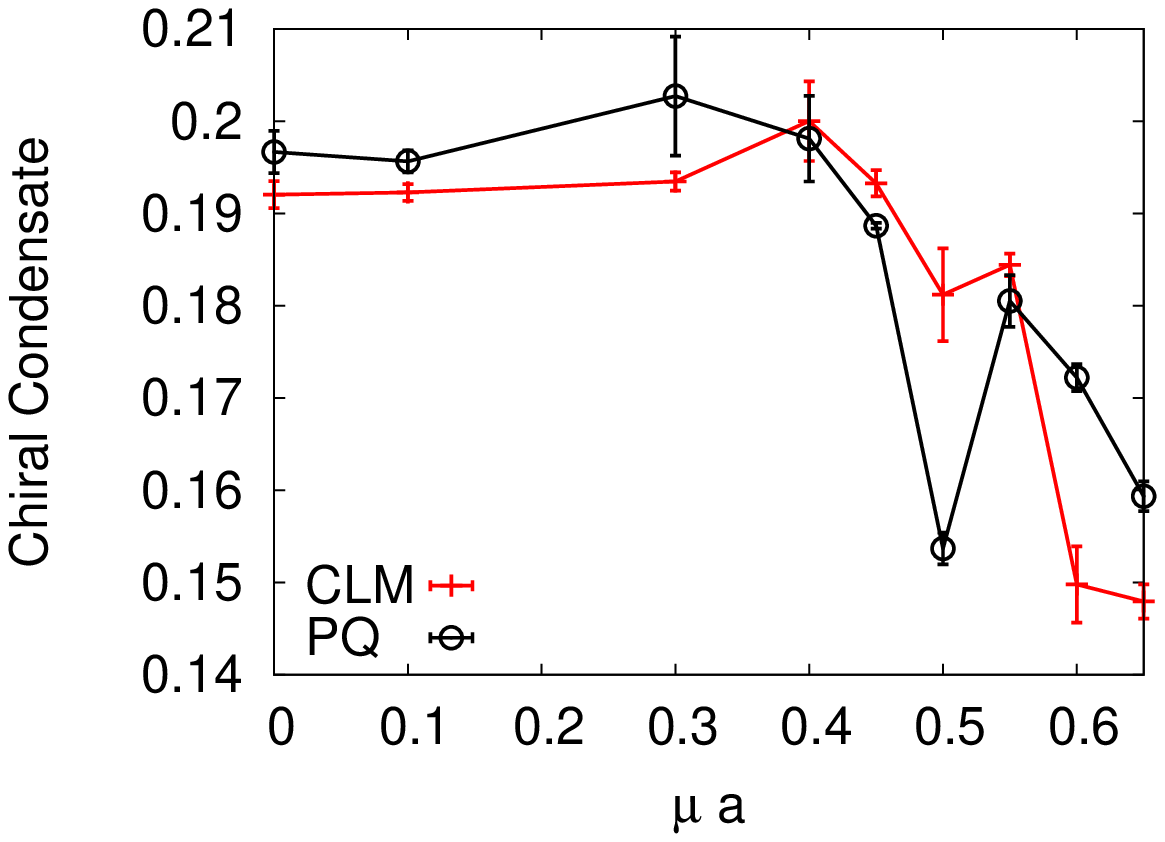}
\caption{The baryon number density (Left) and 
the chiral condensate (Right) are plotted against $\mu a$. 
The symbol referred to as ``CLM'' represents 
the result of QCD with the CLM, 
while the symbol referred to as ``PQ'' represents 
the result of the phase quenched QCD 
with the real Langevin method.}
\label{Fig:quarknumber}
\end{center}
\end{figure}


Figure~\ref{Fig:quarknumber} shows the $\mu$ dependence of
the baryon number density and the chiral condensate 
obtained from the complex Langevin simulation of QCD.
At small $\mu$, the baryon number density is zero and 
the chiral condensate is nonzero.
As we increase $\mu$, the baryon number density 
starts to increase and the chiral condensate starts to decrease
at some point.
At $\mu a = 0.4 \sim 0.5$, the baryon number 
density reaches the order of $0.01$, 
which roughly corresponds to the density for having
one baryon on the lattice, i.e., $1/ N_s^3 =1/4^3 \sim 0.016$.


In Fig.~\ref{Fig:quarknumber} we also plot the results of 
the real Langevin simulation for the phase quenched QCD (PQ).
We observe clear tendency that
the onset value of $\mu$ at which the baryon number 
density starts to increase is smaller for PQ.
It is known that in PQ, the onset value
is $\mu = m_\pi/2$~\cite{Son:2000by},
which can be estimated as $0.265$ 
by using the results \cite{Barbour:1986jf} 
of the mean field analysis for $m a= 0.05$.
In the case of QCD, on the other hand,
the mean field analysis \cite{Barbour:1986jf}
yields an estimate of the onset value 
$\mu a = 0.55$.
Our results are in reasonable agreement with these
predictions considering
finite size effects, which smoothen the transition.


\begin{figure}[htbp] 
\begin{center}
\includegraphics[width=7cm]{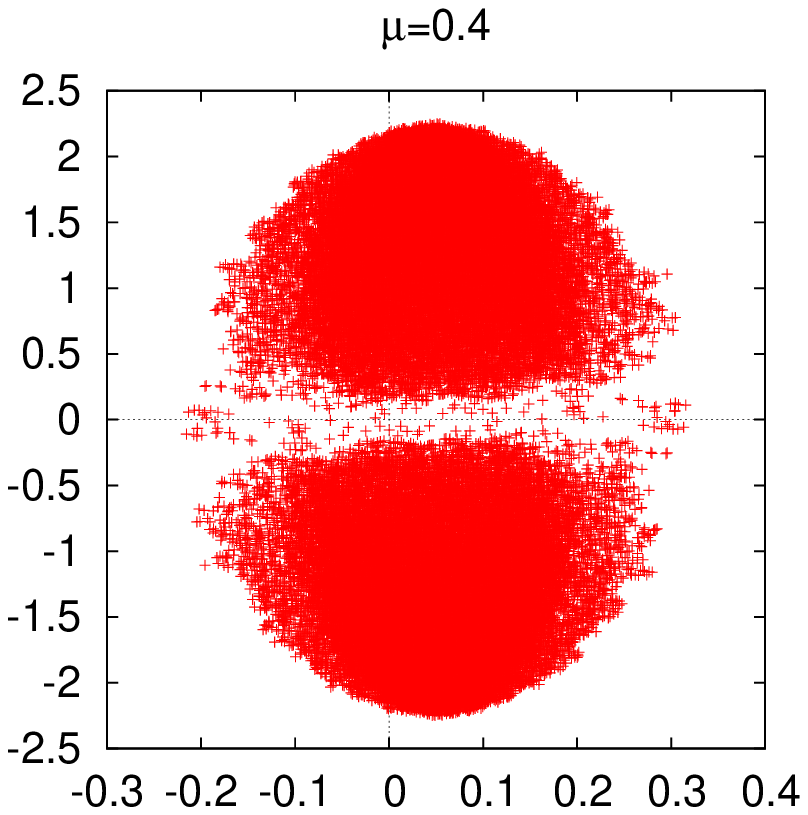}
\includegraphics[width=7cm]{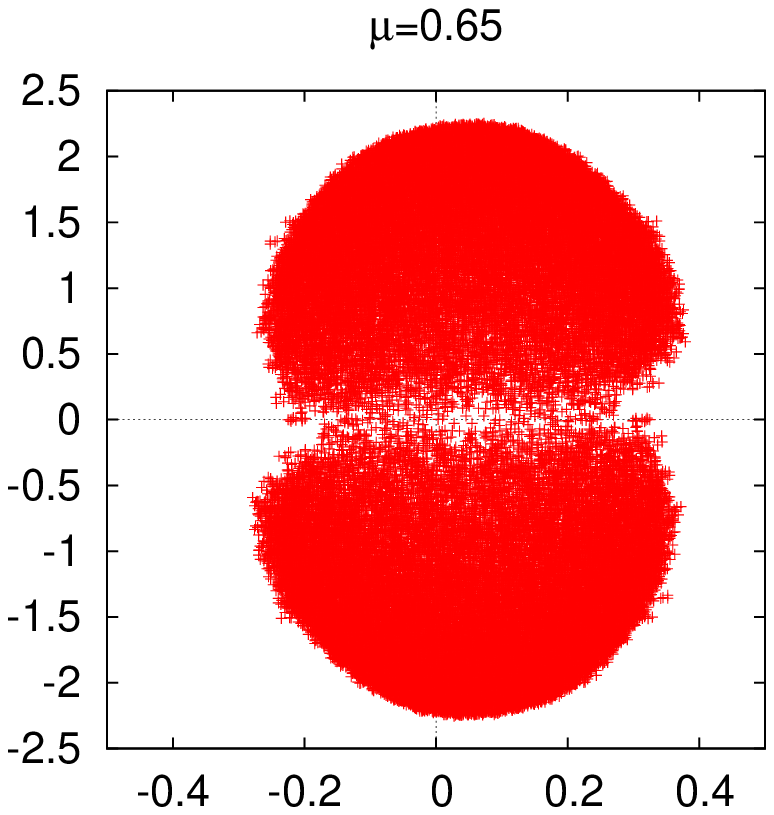}
\caption{Scatter plots of the eigenvalues of the fermion matrix
$M$ for $\mu a = 0.4$ (Left) 
and 0.65 (Right).}\label{Fig:scat_Dev}
\end{center}
\end{figure}

The eigenvalue distribution of the fermion matrix $M$ is 
shown in Fig.~\ref{Fig:scat_Dev}
for $\mu a = 0.4$ and $\mu a = 0.65$.
Let us recall that 
the eigenvalues close to the origin make the fermionic drift term 
large and cause the singular drift problem, which invalidates the CLM.
In fact, we observe a gap along the real axis 
at $\mu=0$ due to finite size effects,
although the lattice setup used in this work corresponds 
to the confined phase.
As we increase $\mu$, 
the eigenvalue distribution extends in the real direction
and the gap becomes smaller.
At $\mu a=0.4$, the eigenvalue distribution still has a gap,
which implies that there are not so many eigenvalues close to the origin.
At $\mu a = 0.65$, however, the gap is not clear any more,
and there are quite a few eigenvalues close to the origin.
At even larger $\mu$, the gap disappears eventually.
Therefore, a crucial question that arises is: up to which value of $\mu$ 
our complex Langevin simulation is valid.



\begin{figure}[htbp]
\begin{center}
\includegraphics[width=7cm]{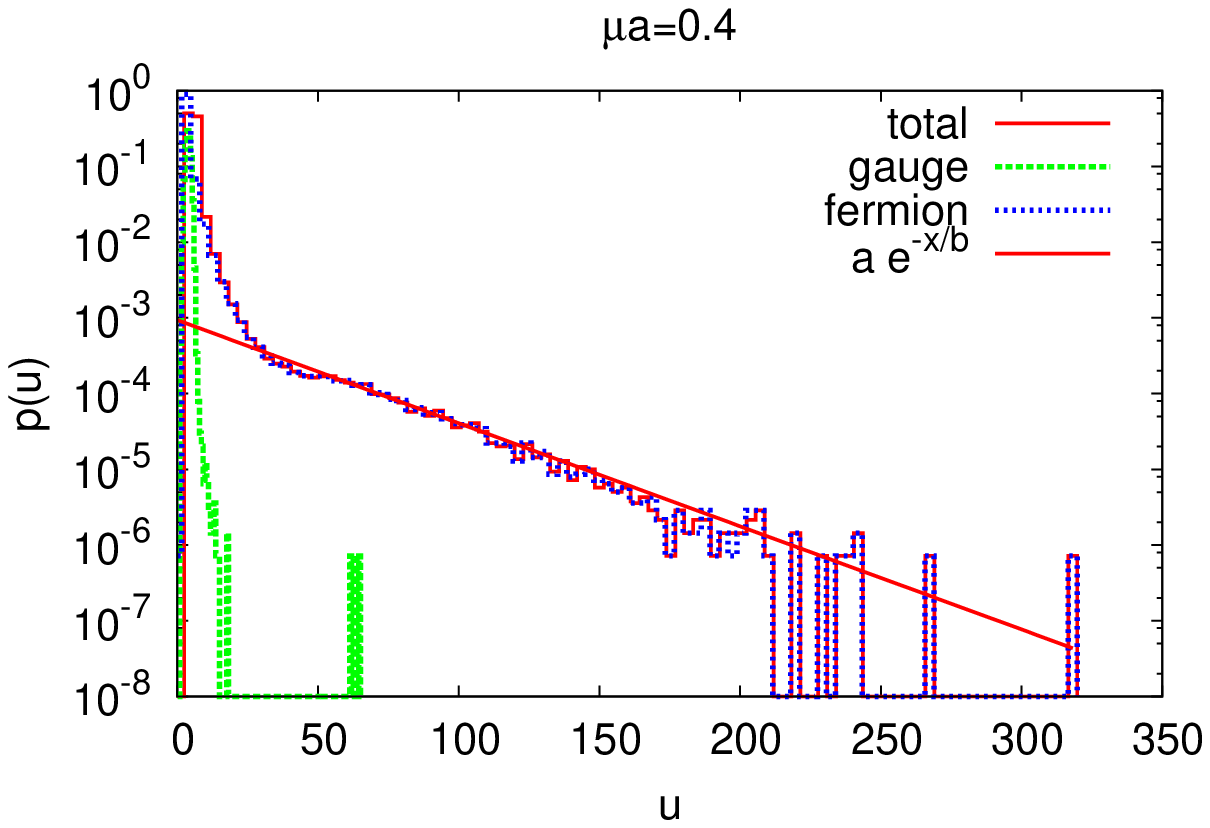}
\includegraphics[width=7cm]{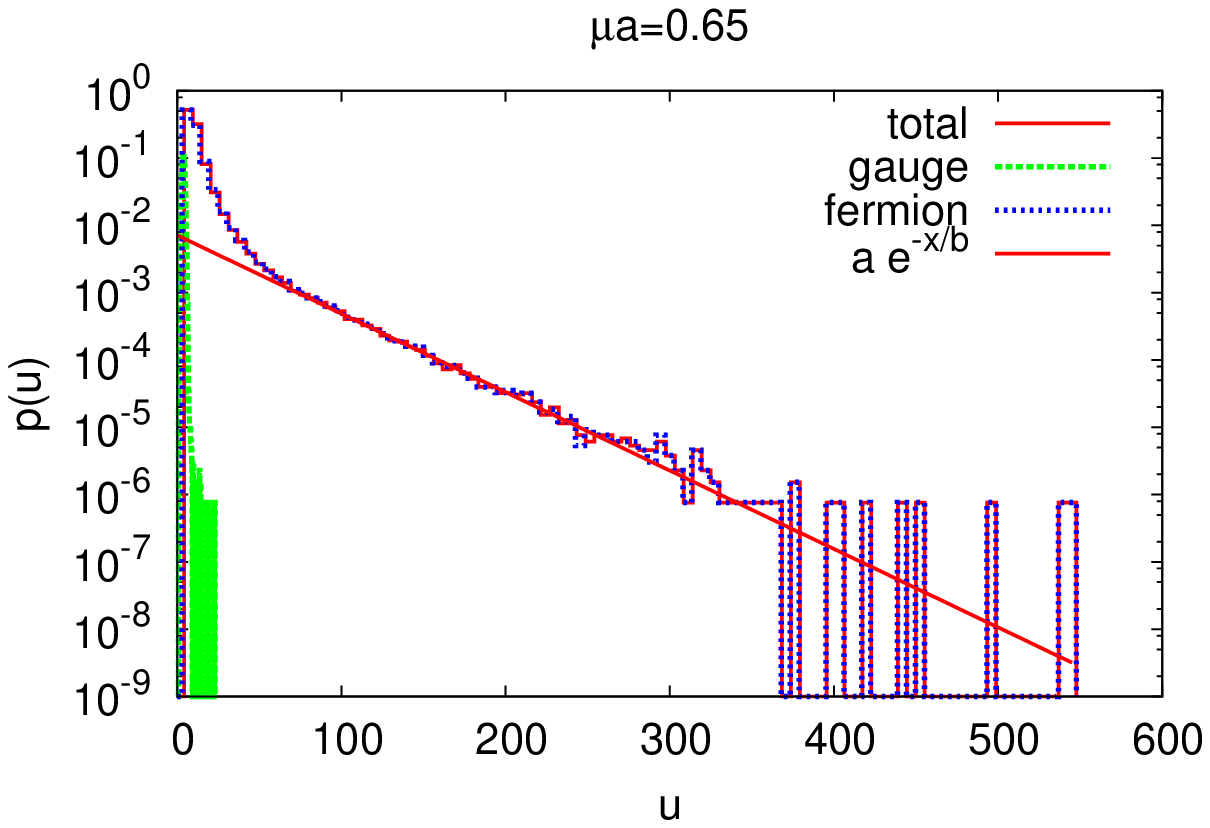}
\caption{The probability distribution of the drift terms 
in the log scale. 
Straight lines represent fits to an exponential behavior
$p(u) = a e^{ - u/b}$.}
\label{Fig:semi_log}
\end{center}
\end{figure}

In order to test the validity of our simulations, 
we use the criterion for correct convergence
\cite{Nagata:2016vkn} based on the probability distribution
$p(u)$ of the drift term, which we plot $p(u)$ 
in Fig.~\ref{Fig:semi_log}
for $\mu a=0.4$ and $\mu a=0.65$ in the log scale.
We find that our results can be fitted
by an exponential behavior $p(u) = a e^{-u/b}$,
excluding the isolated peaks observed at large $u$,
which are not statistically significant.
We have performed similar fits for all values of $\mu$, 
and find that the probability distribution is suppressed exponentially
at large $u$ for $\mu a\lesssim 0.65$.
This implies that the condition for correct convergence
\cite{Nagata:2016vkn} is satisfied in this region.


\section{Summary}

We performed complex Langevin simulations 
for lattice QCD at finite density 
in the confined phase on a $4^3\times 8$ lattice. 
The gauge cooling was used only for the unitarity norm in this
exploratory study and it was found
to stabilize the Langevin dynamics as in the case with the
deconfined phase \cite{Sexty:2013ica}.
Comparing QCD and the phase quenched QCD, we find that the 
onset of the baryon number density occurs at larger $\mu$ in QCD
than in the phase quenched QCD.
The onset value for each case agrees reasonably with 
the predictions of the mean field analysis.
We have also measured the probability distribution of the drift term,
and confirmed that the criterion for correct convergence proposed 
recently is satisfied for $\mu a\lesssim 0.65$, which covers the
interesting region in the present setup.

While our preliminary results are certainly encouraging,
we also find that the auto-correlation time 
at $\mu a \gtrsim 0.4$
is quite long, 
which seems to be related the possible occurrence of the 
singular drift problem at $\mu a \gtrsim 0.7$.
Furthermore, for a larger lattice, it is anticipated that
the gap in the eigenvalue distribution of the fermion matrix
disappears, which will cause a serious singular drift problem.
We are currently trying to develop new techniques to
overcome these problems.

\acknowledgments
This work was supported by Grant-in-Aid for Scientific Research 
(No.~26800154 for K.N.\ and No.~23244057, 
16H03988 for J.N.) from Japan Society for the Promotion of Science.
Simulations were mainly carried out on SX-ACE at RCNP Osaka University 
and CRAY XC40 at YITP Kyoto University. 
SR16000 at KEK was also used for developing a parallel code.
S.S.\ was supported by the MEXT-Supported Program 
for the Strategic Research Foundation at
Private Universities ``Topological Science'' (Grant No.~S1511006).

\providecommand{\href}[2]{#2}\begingroup\raggedright\endgroup



\end{document}